\documentstyle[12pt]{article}
\begin{document}
\title{Stationary Strings and 2-D Black Holes}
\author{V.P. Frolov\thanks{CIAR Cosmology Program;
Theoretical Physics Institute, University of Alberta,
Edmonton, Canada T6G 2J1} and
A.L. Larsen\thanks{Observatoire de Paris,
DEMIRM. Laboratoire Associ\'{e} au CNRS
UA 336, Observatoire de Paris et
\'{E}cole Normale Sup\'{e}rieure. 61, Avenue
de l'Observatoire, 75014 Paris, France.}}
\maketitle
\begin{abstract}
\hspace*{-6mm}A general
description of string excitations in stationary spacetimes
is
developed. If a stationary string passes through
the ergosphere of a 4-dimensional
black hole, its world-sheet describes
a 2-dimensional black (or white) hole with horizon coinciding
with the static
limit of the 4-dimensional black hole. Mathematical results for
2-dimensional black holes can therefore be applied to physical
objects
(say) cosmic strings in the vicinity of Kerr black holes. An
immediate
general result is that the string modes are thermally excited.
The string excitations
are determined by a coupled system of scalar field equations in the
world-sheet metric. In the special case of excitations propagating
along a
stationary string in the equatorial plane of the Kerr-Newman black
hole, they
reduce to the $s$-wave scalar field equations in the 4-dimensional
Reissner-Nordstr\"{o}m black hole. We briefly discuss possible
applications of
our results  to the black hole information puzzle.
\end{abstract}
The
interaction of strings and black holes is an interesting subject and
many publications appeared recently, which discuss different aspects
of
this problem. The aim of this paper is to attract the attention to
the
remarkable fact, that the study of propagation of perturbations along
a
stationary string located in the gravitational field of a stationary
4-dimensional black hole, is directly related with physics of
2-dimensional
black holes.
If one neglects the gravitational effects of a string and
assumes that its thickness is zero, then a string configuration in a
given
gravitational field is  a minimal surface, and its equations of
motion
can be obtained
by variation of the Goto-Nambu action. An important case is when both
the
gravitational field and a string configuration are stationary. This
case
corresponds to a physical situation of a string which is in
equilibrium
in the gravitational field.
Geometrically
the equilibrium configuration is characterized by the property that
the
stationary Killing vector $\xi$ is tangent everywhere to the string
surface
$\Sigma$. The problem of finding the equilibrium string
configurations in a
stationary spacetime can be reduced to the problem of solving
geodesic
equations in a 3-dimensional unphysical
space \cite{fro}. The remarkable property
of the gravitational field of a stationary black hole is that these
geodesic
equations can be solved in quadratures \cite{fro,car1,hei}.
Consider a 2-dimensional  time-like world-sheet $\Sigma$ of
the string defined by the equations  $x^{\mu}=x^{\mu}(\zeta^{A})$
($x^{\mu}$ ($\mu =0,...,4$) denote  spacetime coordinates and
$\zeta^{A}$ ($A=0,1$) are the coordinates on the world-sheet).
The metric  $G_{AB}$  induced on $\Sigma$ reads:
\begin{equation}
G_{AB}=g_{\mu \nu}x^{\mu}_{,A}x^{\nu}_{,B} .\label{1}
\end{equation}
where $g_{\mu\nu}$ is the spacetime metric. Consider a stationary
spacetime and
let $\xi^{\mu}$ be the corresponding
4-dimensional Killing vector. We call a string stationary if
$\xi^{\mu}$ is
tangent to its world sheet $\Sigma$. Denote by $\eta^{A}$ a
2-dimensional
vector $\eta^{A}=G^{AB}x^{\mu}_{,B}\xi_{\mu}$. It is easy to show
that:
\begin{equation}
\xi^{\mu}=\eta^{A}x^{\mu}_{,A},
\end{equation}
\begin{equation}
\eta_{A|B}=x^{\mu}_{,A}x^{\nu}_{,B}\xi_{\mu;\nu} ,
\end{equation}
where semicolon and vertical line denote covariant derivatives
with respect to 4 and 2-dimensional metrics, respectively. The last
relation implies that $\eta$ is a Killing
vector for the induced metric $G_{AB}$.
Denote:
\begin{equation}
F=-g_{\mu\nu}\xi^{\mu}\xi^{\nu}
=-G_{AB}\eta^{A}\eta^{B}.
\end{equation}
Then the
induced line-element $dl^2=G_{AB}d\zeta^{A}
d\zeta^{B}$ on $\Sigma$ can be written as:
\begin{equation}
dl^2=-Fd\hat{\tau}^2+F^{-1}d\hat{\sigma}^2,\label{2}
\end{equation}
where $\hat{\tau}=\hat{\tau}(\zeta^{A}),\;\;
\hat{\sigma}=\hat{\sigma}(\zeta^{A}).$
This representation is valid in the regions where $F\ne 0$, and hence
$\xi$ is
either time-like or space-like.
We assume that the 4-dimensional spacetime is asymptotically flat and
contains
a black hole. We assume also that
$\hat{\sigma}=\infty$ corresponds to the points of the string located
in the
asymptotically flat region of the physical spacetime,
so that $F(\hat{\sigma}=\infty)=1$. The metric, Eq.(5),
describes a 2-dimensional black hole if $F=0$ at finite value of
$\hat{\sigma}$. At this
point the Killing 2-vector $\eta$ is null. It happens at the points
where the
string world-sheet crosses the infinite-red-shift surface
(the static limit), i.e. the surface where $\xi^2=0$. For a static
black hole
this surface coincides with the event horizon. For stationary
(rotating) black holes it is lying outside the horizon.
The region located between infinite-red-shift surface and
the event
horizon of a stationary black hole is known as the ergosphere.
Points of the string
located inside the ergosphere thus
correspond to the interior of the 2-dimensional
black hole.

The main observation we would like to make now is that perturbations
propagating along the string can be described by a coupled system of
a pair of scalar field equations in the 2-dimensional metric
$G_{AB}$ \cite{all}.
For this reason, if a stationary string is passing through the
ergosphere
of a 4-dimensional black hole, the physics of string excitations is
effectively
reduced to  the physics of 2-dimensional black holes.

We shall first derive the differential equations describing the
perturbations
propagating along a stationary string configuration embedded in an
arbitrary
stationary 4-dimensional spacetime. We then consider, as a special
example,
stationary strings in the background of a Kerr-Newman black hole.
\vskip 12pt
\hspace*{-6mm}To be more specific, we write the metric of a generic
4-dimensional
stationary spacetime in the form:
\begin{equation}
g_{\mu\nu}=\left( \begin{array}{cc} -F & -FA_{i}\\
-FA_{i} & -FA_{i}A_{j}+H_{ij}/F\end{array}\right),
\end{equation}
where $\partial_t F=0,\;\partial_t A_{i}=0,\;\partial_t H_{ij}=0.$
That is to say,
the Killing vector is given explicitly by:
\begin{equation}
\xi^\mu=(1,\;0,\;0,\;0),\;\;\;\;\;\;\xi_\mu=(-F,\;-FA_{i}),
\end{equation}
consistent with the notation of Eq.(4).
A stationary string configuration is parametrized in the following
way:
\begin{equation}
t=x^0=\tau,\;\;\;\;\;x^{i}=x^{i}(\sigma){\;};\;\;\;\;\;\;\;\;
(\tau\equiv\zeta^0,\;\sigma\equiv\zeta^1).
\end{equation}
Then the equations of motion corresponding to the Goto-Nambu action
reduce
to
\cite{fro}:
\begin{equation}
x^{i}{''}+\tilde{\Gamma}^{i}_{jk}x^{j}{'}
x^{k}{'}=0,\;\;\;\;\;\;H_{ij}x^{i}{'}x^{j}{'}=1,
\end{equation}
where $\tilde{\Gamma}^{i}_{jk}$ is the Christoffel connection for the
metric $H_{ij}$ and a prime denotes differentiation with respect to
$\sigma.$ The induced metric on the world-sheet now takes the form:
\begin{equation}
G_{AB}=\left( \begin{array}{cc} -F & -FA\\
-FA & -FA^2+1/F\end{array}\right);\;\;\;\;\;\;\;\;A\equiv
A_{i}x^{i}{'},
\end{equation}
so that $\mbox{det}\ G=-1$.
The following coordinate transformation on the world-sheet:
\begin{equation}
\hat{\tau}=\tau+\int^{\sigma}A\;d\sigma,\;\;\;\;\;\;\;\;\hat{\sigma}=\sigma,
\end{equation}
brings the induced line element into the form of Eq.(5), that is:
\begin{equation}
\hat{G}_{AB}=\left( \begin{array}{cc} -F & 0\\
0 & 1/F\end{array}\right).
\end{equation}

A covariant approach describing the propagation of perturbations
along
an arbitrary string configuration embedded in an arbitrary
spacetime, was developed by the present authors in Ref.\cite{all}
(see also
\cite{car2,guv}). The
general transverse (physical) perturbation around a background
Goto-Nambu string configuration is written as:
\begin{equation}
\delta x^\mu=\Phi^R n^\mu_R,\;\;\;\;\;\;(R=2,3),
\end{equation}
where the two vectors $n^\mu_R,\;$ normal to the string world-sheet,
fulfill:
\begin{equation}
g_{\mu\nu}n^\mu_R n^\nu_S=\delta_{RS},\;\;\;\;\;\;\;\;\;\;g_{\mu\nu}
x^\mu_{,A}n^\nu_R=0,
\end{equation}
as well as the completeness relation:
\begin{equation}
g^{\mu\nu}=G^{AB}x^\mu_{,A} x^\nu_{,B}+\delta^{RS}n^\mu_R n^\nu_S.
\end{equation}
It can be shown \cite{all} that the perturbations $\Phi^R$
are determined by the following effective action:
\begin{equation}
{\cal S}_{\mbox{eff.}}=\int_{}^{}
d^2\zeta\sqrt{-G}\;\Phi^R\left\{G^{AB}
(\delta^T_R\nabla_A+\mu_R\hspace*{1mm}^T\hspace*{1mm}_A)
(\delta_{TS}\nabla_B+\mu_{TSB})+
{\cal V}_{RS}\right\}\Phi^S,
\end{equation}
where ${\cal V}_{RS}={\cal V}_{(RS)}$ are scalar potentials
and $\mu_{RS}\hspace*{1mm}^A=\mu_{[RS]}\hspace*{1mm}^A$
are vector potentials on
$\Sigma,$
determined by its embedding into the 4-dimensional spacetime,
and  $\nabla_A$ is the covariant derivative with
respect to  the metric $G_{AB}$.  The vector potentials
$\mu_{RSA}$
coincide with  the normal fundamental form:
\begin{equation}
\mu_{RSA}=g_{\mu\nu}n^\mu_R x^\rho_{,A}\nabla_\rho n^\nu_S,
\end{equation}
where $\nabla_\rho$ is the covariant derivative with respect to
the metric $g_{\mu\nu}$. The scalar potentials ${\cal V}_{RS}$
are defined as:
\begin{equation}
{\cal V}_{RS}\equiv\Omega_{RAB}\Omega_S\hspace*{1mm}^{AB}-
G^{AB}x^\mu_{,A}x^\nu_{,B}R_{\mu\rho\sigma\nu}n^\rho_R
n^\sigma_S,
\end{equation}
where $\Omega_{RAB}$ is the second fundamental form:
\begin{equation}
\Omega_{RAB}=g_{\mu\nu}n^\mu_R x^\rho_{,A}\nabla_\rho x^\nu_{,B},
\end{equation}
and $R_{\mu\rho\sigma\nu}$ is the Riemann tensor corresponding to
the metric $g_{\mu\nu}$.
The equation describing the propagation of  perturbations along the
string world-sheet is:
\begin{equation}
\left\{ G^{AB}(
\delta^T_R {\nabla}_A+{\mu}_R\hspace*{1mm}^T
\hspace*{1mm}_A)(
\delta_{TS}{\nabla}_B+{\mu}_{TSB})
+{{\cal V}}_{RS}\right\}\Phi^S=0.
\end{equation}

We obtain now explicit expressions for vector
${\mu}_{RS}\hspace*{1mm}^A$ and
scalar ${\cal V}_{RS}$ potentials which enter the propagation
equations for
a stationary string obeying Eqs.(9), in the background (6).
Using Eqs.(14), the normal vectors $n^\mu_R$ take the form:
\begin{equation}
n^\mu_R=(-A_{i}n^{i}_R,\;n^{i}_R),\;\;\;\;\;\;n_{\mu
R}=(0,\;F^{-1}H_{ij}
n^{j}_R)\equiv (0,\;n_{iR}).
\end{equation}
It is convenient to introduce also  3-dimensional vectors
\begin{equation}
\tilde{n}^{i}_R=|F|^{-1/2}n^{i}_R,
\end{equation}
which together with $x^{i}{'}$
form an  orthonormal
system $(x^{i}{'},\tilde{n}^{i}_2,\tilde{n}^{i}_3)$ in the
3-dimensional
space with metric $H_{ij}$.
We note that there is an ambiguity
$\tilde{n}^{i}_R\rightarrow \tilde{n}^{i}_R+\delta(\tilde{n}^{i}_R)$
in the choice of pair of normal
vectors  $\tilde{n}^{i}_R:$
\begin{equation}
\delta(\tilde{n}^{i}_R)=
\Lambda_R\hspace*{1mm}^S\tilde{n}^{i}_S\;
;\;\;\;\;\;\;\Lambda_{RS}(\zeta)=-\Lambda_{SR}(\zeta).
\end{equation}
Under these 'gauge'  transformations, which leave
invariant the perturbations (13) and the effective action (16),
the quantities $\Phi^S$, ${\mu}_{RSA}$, and ${\cal V}_{RS}$
are transformed as follows:
\begin{eqnarray}
\delta(\Phi^R)\hspace*{-2mm}&=&\hspace*{-2mm}
\Lambda^R\hspace*{1mm}_S\Phi^S,\nonumber\\
\delta(\mu_{RSA})
\hspace*{-2mm}&=&\hspace*{-2mm}\Lambda_R\hspace*{1mm}^T\mu_{TSA}
-\Lambda_S\hspace*{1mm}^T\mu_{TRA}-\Lambda_{RS,A},\\
\delta({{\cal V}}_{RS})\hspace*{-2mm}&=&\hspace*{-2mm}
\Lambda_R\hspace*{1mm}^T{{\cal V}}_{TS}+
\Lambda_S\hspace*{1mm}^T{{\cal V}}_{RT} .\nonumber
\end{eqnarray}
We fix
the freedom of  'gauge'  by
choosing the vectors $\tilde{n}^{i}_R$ to be covariantly constant in
the 3-dimensional  space \cite{all}.
After straightforward but tedious
calculations we get:
\begin{equation}\label{mu}
\mu_{RS}\hspace*{2mm}^{A}=\eta^A\;\frac{A_{ij}}{2}
n^{i}_R n^{j}_S=\eta^{A}\;\frac{1}{2F^2}\xi_{[0,i}\xi_{j]}
n^{i}_R n^{j}_S,
\end{equation}
\begin{eqnarray}
{\cal
V}_{RS}\hspace*{-2mm}&=&\hspace*{-2mm}-x^{i}{'}x^{j}{'}\tilde{R}_{iklj}
n^{k}_R
n^{l}_S+\frac{\delta_{RS}}{2}x^{i}{'}x^{j}{'}(\tilde{\nabla}_{i}
\tilde{\nabla}_{j}F)\nonumber\\
\hspace*{-2mm}&-&\hspace*{-2mm}\frac{\delta_{RS}}{4F}x^{i}
{'}x^{j}{'}(\tilde{\nabla}_{i}F)(\tilde{\nabla}_{j}F)
+\frac{F}{4}\delta^{TU}A_{il}A_{kj}n^{i}_U n^{j}_T n^{k}_R n^{l}_S.
\end{eqnarray}
Here the "field-strength" $A_{ij}$ is defined by:
\begin{equation}
A_{ij}=A_{i,j}-A_{j,i}=\tilde{\nabla}_{j}A_{i}-\tilde{\nabla}_{i}A_{j},
\end{equation}
where $\tilde{\nabla}_{i}$ is the covariant derivative with respect
to the
metric $H_{ij}$
and $\tilde{R}_{iklj}$ is the Riemann tensor corresponding to the
metric $H_{ij}.$ It is easy to verify that for our choice of 'gauge'
the
vector potentials $\mu_{RS}\hspace*{2mm}^{A}$ obey the analog
of the Lorentz
gauge conditions  $\nabla_A\mu_{RS}\hspace*{1mm}^{A}=0$.
The anti-symmetric products of normal vectors, appearing in the
scalar and vector potentials, Eqs.(25)-(26),
can be eliminated using the identity:
\begin{equation}
\epsilon^{RS}\tilde{n}^{i}_R\tilde{n}^{j}_S=\tilde{e}^{ijk}H_{kl}x^{l}
{'};
\;\;\;\;\;\;\;\;\tilde{e}^{ijk}=(H)^{-1/2}\epsilon^{ijk}.
\end{equation}
In particular we have:
\begin{equation}
x^{i}{'}x^{j}{'}\tilde{R}_{iklj}n^{k}_R n^{l}_S=
F\delta_{RS}(\frac{1}{2}\tilde{R}-\tilde{R}_{ij}x^{i}{'}
x^{j}{'})-\tilde{R}_{ij}n^{i}_R n^{j}_S,
\end{equation}
\begin{equation}
\frac{F}{4}\delta^{TU}A_{il}A_{kj}n^{i}_U n^{j}_T n^{k}_R n^{l}_S=
\frac{F^3}{4}\delta_{RS}(A_{ik}A_{j}\;^{k}x^{i}{'}x^{j}{'}-\frac{1}{2}
A_{ij}
A^{ij}).
\end{equation}
By using the equations (29)-(30) we get:
\begin{equation}
\mu_{RS}\hspace{1mm}^A=\mu \epsilon_{RS}\eta^A\;\; ;\;\;
\;\;\;\;\;\ \mu=\frac{F}{4}A_{ij}
\tilde{e}^{ijk}H_{kl}x^{l}{'},
\end{equation}
\begin{eqnarray}
{\cal  V}_{RS}\hspace{-2mm}&=&\hspace{-2mm}\delta_{RS}
[\frac{x^{i}{'}x^{j}{'}}{2}(\tilde{\nabla}_{i}\tilde{\nabla}_{j}F)-
\frac{1}{4F}x^{i}
{'}x^{j}{'}(\tilde{\nabla}_{i}F)(\tilde{\nabla}_{j}F)-
F(\frac{1}{2}\tilde{R}-\tilde{R}_{ij}x^{i}{'}
x^{j}{'})\nonumber\\
\hspace{-2mm}&+&\hspace{-2mm}
\frac{F^3}{4}\delta_{RS}(A_{ik}A_{j}\;^{k}x^{i}{'}x^{j}{'}-\frac{1}{2}
A_{ij}A^{ij})]+\tilde{R}_{ij}n^{i}_R n^{j}_S .
\end{eqnarray}

As a special application of the above formalism, we now
consider
perturbations propagating
along a stationary string in the Kerr-Newman black hole
background. The Kerr-Newman metric in  Boyer-Lindquist coordinates
\cite{boy} reads:
\begin{equation}\label{BL}
ds^2=-\frac{\Delta}{\rho^2} [dt -a\sin^2\theta d\phi ]^2
+\frac{\sin^2\theta}{\rho^2}
[(r^2+a^2)d\phi -a dt]^2 +\frac{\rho^2}{\Delta}dr^2 +\rho^2 d\theta^2
,
\end{equation}
where  $\Delta=r^2-2Mr+Q^2+a^2$ and $\rho^2=r^2+a^2\cos^2\theta$.
This metric  is of the form  (6) with:
\begin{eqnarray}
H_{rr}=\frac{\Delta-a^2\sin^2\theta}{\Delta},
\hspace{.6cm}
H_{\theta\theta}=\Delta-a^2\sin^2\theta,
\hspace{.6cm}
H_{\phi\phi}=\Delta\sin^2\theta,
\end{eqnarray}
\begin{equation}
F=\frac{\Delta-a^2\sin^2\theta}{r^2+a^2\cos^2\theta},
\hspace{.6cm}
A_\phi=a\sin^2\theta\;\frac{2Mr-Q^2}{\Delta-a^2\sin^2\theta}.
\end{equation}
The unperturbed stationary string configurations are obtained by
solving
Eqs.(9), \cite{fro}:
\begin{eqnarray}\label{30}
(H_{rr}r')^2\hspace*{-2mm}&=&\hspace*{-2mm}\frac{a^2 b^2}{\Delta^2}-
\frac{q^2}{\Delta}+1,\nonumber\\
(H_{\theta\theta}\theta')^2\hspace*{-2mm}&=&\hspace*{-2mm}q^2-
\frac{b^2}{\sin^2\theta}-a^2\sin^2\theta \equiv q^2-U(\theta),\\
(H_{\phi\phi}\phi')^2\hspace*{-2mm}&=&\hspace*{-2mm}b^2,\nonumber
\end{eqnarray}
where $b$ and $q$ are integration constants. Equations (\ref{30}) are
evidently invariant with respect to the reflection $\phi\rightarrow
-\phi$.
It means that if $(r(\sigma), \theta(\sigma), \phi (\sigma) )$ is
a
solution of Eqs.(\ref{30}) then $(r(\sigma), \theta(\sigma), -\phi
(\sigma) )$
is also a solution.

We  consider at first the case when a stationary
string is located in the equatorial plane $\theta=\pi/2,$
corresponding to $|b|\geq |a|$ and $q^2=a^2+b^2.$
In that case, the two
covariantly constant normal vectors $\tilde{n}^{i}_R,$ introduced in
Eq.(22), are given by:
\begin{equation}
\tilde{n}^{i}_2\equiv\tilde{n}^{i}_\perp=\frac{1}{\sqrt{|r^2-2Mr+Q^2|}}
(0,\;1,\;0),
\end{equation}
\begin{equation}
\tilde{n}^{i}_3\equiv\tilde{n}^{i}_\parallel=\frac{1}{\sqrt{|r^2-2Mr+Q
^2|}}
(-b,\;0,\;H_{rr}r').
\end{equation}
It is now straightforward to compute the vector and scalar potentials
given by Eqs.(31),(32):
\begin{equation}
\mu=0,
\end{equation}
\begin{eqnarray}
{\cal V}_{\perp\perp}\hspace*{-2mm}&=&\hspace*{-2mm}
{\cal
V}_{\parallel\hspace*{1mm}\parallel}+\frac{2(M^2-Q^2)(a^2-b^2)}
{r^2(\Delta-a^2)^2}\nonumber\\
\hspace*{-2mm}&=&\hspace*{-2mm}\frac{M^2-Q^2}{r^2(\Delta-a^2)}
[1+\frac{2(a^2-b^2)}{\Delta-a^2}]+\frac{(r-M)(b^2-a^2)}{(\Delta-a^2)^2}
[\frac{M}{r^2}-\frac{Q^2}{r^3}]\nonumber\\
\hspace*{-2mm}&+&\hspace*{-2mm}\frac{\Delta-b^2}{(\Delta-a^2)^2}
[-\frac{2M}{r}+\frac{3Q^2}{r^2}+\frac{3M^2}{r^2}-\frac{6MQ^2}{r^3}
+\frac{2Q^4}{r^4}].
\end{eqnarray}

In the generic case, $|b|>|a|,$ the stationary configuration
in the equatorial plane describes
an infinitely long open string with two "arms" in the
asymptotically flat regions and a turning point {\it outside} the
ergosphere
\cite{fro}. The perturbation equations (20) with the potentials
(39),(40)
then determine the reflection and transmission of waves (`phonons')
travelling along the string between the
asymptotically flat regions \cite{fro,all2}. However, in
the special case $|b|=|a|,$ the string passes the static
limit and spirals inside the ergosphere towards the horizon. This is
the case
we are interested in, c.f. the discussion after Eq.(5).
Notice that the potential (40), in general being divergent at the
static limit $(\Delta=a^2),$ is finite for $|b|=|a|,$ i.e. the
divergences precisely cancel out in the particular case where the
string actually crosses the static limit.

Let us now consider the case $|b|=|a|$ in more detail.
We use the notations $\Sigma_{\pm}$ for a pair of
string configurations connected by the reflection
transformation, $\phi\rightarrow
-\phi,$ discussed after Eq.(36).
In order to make this prescription unique we choose
$dr/d\phi >0$ for $\Sigma_+$ and $dr/d\phi <0$ for $\Sigma_-$.
Then we find from Eq.(11):
\begin{equation}
\hat{\sigma}=r,\;\;\;\;\;\;\;\;\;\;\;\hat{\tau}=t\pm
a^2\int^r\;\frac{2Mr
-Q^2}
{\Delta(\Delta-a^2)}\;dr,
\end{equation}
where the signs $\pm$ correspond to the string configuration
$\Sigma_{\pm}$. For both  configurations $\Sigma_{\pm}:$
\begin{equation}
\hspace*{-6mm}F=1-\frac{2M}{r}+\frac{Q^2}{r^2},
\end{equation}
thus the world-sheet line element (5) takes the form of a
2-dimensional
Reissner-Nordstr\"{o}m black hole. The parameters of mass $M$
and charge $Q$ of this 2-dimensional black hole are the same
as the corresponding  parameters of the original 4-dimensional
Kerr-Newman
metric.  The world-sheet spatial   coordinate
$\hat{\sigma}$ equals the Boyer-Lindquist radial coordinate $r,$
while
the
world-sheet time $\hat{\tau}$ approaches the Boyer-Lindquist time $t$
in the
asymptotically flat region, $r\rightarrow\infty.$  The coordinates
$(\hat{\tau}, r)$ cover only the exterior region of the string hole.
One can easily obtain the string metric valid in a wider region.
For this purpose it is convenient to introduce  the
Eddington-Finkelstein coordinates
$(u_{\pm},\varphi_{\pm}):$
\begin{equation}\label{EF}
dt=du_{\pm}\pm \Delta^{-1} (r^2+a^2)dr,\hspace{1cm}
d\phi=d\varphi_{\pm}\pm \Delta^{-1}a dr ,
\end{equation}
and  to rewrite the Boyer-Lindquist metric (\ref{BL}) as \cite{MTW}:
\begin{eqnarray}\label{EF1}
ds^2\hspace*{-2mm}&=&\hspace*{-2mm}
-\frac{\Delta}{\rho^2}[du_{\pm} -a\sin^2\theta d\varphi
_{\pm}]^2
+\frac{\sin^2\theta}{\rho^2}
[(r^2+a^2)d\varphi_{\pm} -a du_{\pm}]^2\nonumber \\
\hspace*{-2mm}&+&\hspace*{-2mm}
\rho^2 d\theta^2 \mp 2dr [du_{\pm} -a\sin^2\theta d\varphi_{\pm} ].
\end{eqnarray}
In these coordinates, the strings $\Sigma_{\pm}$ are described by
equations $\theta=\pi/2$, $\varphi_{\pm}=\mbox{const.},$ so that the
induced metric on $\Sigma_{\pm}$ is:
\begin{equation}\label{2EF}
ds^2=-Fdu_{\pm}^2\mp 2dr du_{\pm}.
\end{equation}
This metric  for $\Sigma_+$ describes a black hole, while  for
$\Sigma_-$
it describes a white hole.
In both cases the perturbation  equations (20) on $\Sigma_{\pm}$
reduce to:
\begin{equation}
\Box\Phi^R=\frac{\Box r}{r}\Phi^R;\;\;\;\;\;\;\;\;R={\tiny{
\perp,\;\parallel}}
\end{equation}
where $\Box$ is the d'Alambertian on the world-sheet,
$\Box=G^{AB}\nabla_A
\nabla_B$. We also have $\Box r= F_{,r}$.

Several interesting remarks are now in order. First notice that the
perturbation equations for the two transverse polarizations $\Phi^R$
are decoupled
and {\it identical}. Secondly, Eq.(46) is precisely the $s$-wave
scalar field equation in the 4-dimensional Reissner-Nordstr\"{o}m
black
hole background:
\begin{equation}
\Box^{(4)}\phi=0,\;\;\;\;\;\;g_{\mu\nu}^{(4)}=
\mbox{diag}(-F,\;1/F,\;r^2,\;r^2\sin^2\theta),
\end{equation}
with $F$ given by Eq.(42).
The decomposition $\phi=\sum r^{-1}\Phi_l (r,t)\;Y_{lm}(\theta,\phi)$
yields:
\begin{equation}
-\frac{1}{F}\partial_t^2\Phi_l+\partial_r(F\partial_r\Phi_l)=
\frac{F_{,r}}{r}\Phi_l+
\frac{l(l+1)}{r^2}\Phi_l,
\end{equation}
which is identical to equation (46) for $l=0,$ as is most easily seen
by writing Eq.(46) in the $(\hat{\tau},r)$ world-sheet coordinate
system.
(Equation (48) has
been studied
in detail in the literature, see for instance \cite{chan} and
references
given therein.)
The string exitations for a stationary string passing through the
ergosphere in the equatorial plane
of a Kerr-Newman black hole are therefore described mathematically
by the $l=0$ scalar waves in the background of a 4-dimensional
Reissner-Nordstr\"{o}m black hole.

The above results allow  generalization to the case when a
stationary string
is located not in the equatorial plane but on the cone
$\theta=\theta_0 \ne \pi/2$.
For this case the parameters which enter the equations (36)
are related as $q^2=2|ab|$. The corresponding $\theta_0$ is
determined
as the minimum of the potential $U(\theta)$ and is, $\sin ^2
\theta_0=|b/a|$.
This relation implies that $|b|<|a|$.
The remarkable fact is that such a string allows a simple geometrical
description. The Kerr-Newman metric possesses two
principal null geodesic congruences, one of them is formed by
incoming and the other  by outgoing principal null rays \cite{MTW}.
Take one of these null geodesics $\gamma_{\pm}$ ($-$ stands for
an outgoing and $+$ stands for an incoming ray). Consider
two-dimensional
surfaces $\Sigma_{\pm}$ formed by Killing trajectories passing
through $\gamma_{\pm}$. It is possible to prove that $\Sigma_{\pm}$
is a minimal surface and it describes an "equilibrium" string
configuration
with the parameter $q^2=2|ab|.$
Two string configurations $\Sigma_{\pm}$
differ by signs of $r'$ in (36). The metric induced on both
surfaces $\Sigma_{\pm}$ possesses the Killing horizon. In case of
$\Sigma_+$ it describes a black hole, while in case of $\Sigma_-$
it describes a white hole (for a white hole a future directed
timelike
curve crossing  the Killing horizon enters the black hole exterior).
The above considered case of a stationary string located
on the equatorial plane and crossing
the infinite-red-shift surface is a special example of stationary
cone strings. The perturbation equations for cone strings will be
considered somewhere else.

To summarize, we have shown that the metric induced on a stationary
string crossing the infinite-red-shift surface  describes a
two-dimensional
geometry of a black or white hole. It opens remarkable possibilities
to  apply results of
mathematical study of two-dimensional black holes to
physical objects  (cosmic strings in the vicinity of a black hole),
which at least in principle allow experimental observations.
In particular, in the presence of the horizon on $\Sigma_+$
(for the two-dimensional string black hole) the conditions that the
quantum state is regular near the horizon implies that
 the string perturbations $\Phi^R$ are to be
thermally excited. It means that there will be a thermal flux of the
string
excitations ('phonons') propagating to infinity which forms the
corresponding
Hawking radiation. For a radial string crossing the event horizon
of the Schwarzschild black hole this effect was considered in
\cite{law}. We would like to stress that the analogous radiation
will
also be present  when a stationary string crosses the
infinite-red-shift surface (which for a rotating black hole is
located outside
the horizon).
String perturbations ('phonons') generated in
the region lying inside the 2-horizon and propagating along the
string $\Sigma_+$
cannot escape to infinity.
But it is well known that the causal signals emitted in the
ergosphere
and propagating in the 4-dimensional space-time can reach a
distant observer. One can use such signals ('photons') in order to
get information from the interior of the two-dimensional string
black  hole.
This situation is similar to one which happens in  a
'dumb' hole considered by Unruh \cite{Unruh}.  In order to define
a 'dumb' hole one uses the causal structure connected with
the propagation of phonons. The photons propagating with
supersonic velocity can escape a 'dumb' hole interior.
The nice feature which differs our model is that it is constructed
in the framework of completely relativistic theory.

Possibility of getting information from a black hole interior was
also
discussed in \cite{FrNo:93}, where  a gedanken experiment was
proposed in which  a traversable wormhole is used.
The mechanism discusses in the present paper which makes it
possible to extract
information from the interior of string black holes is different.
It is connected with the presence of extra-dimensions and does
not require
non-trivial topology.
The possibility of the information extraction from the interior of
a string black hole might give new insight to the problem of
information loss in black holes.
In particular the arguments of  Ref.\cite{Gott:95}  being applied to
string black holes indicate that black hole complementarity
may be inconsistent, at least for these black hole models.
\vskip 12pt
\hspace*{-6mm}{\bf Acknowledgements:}\\
The work of V.P. Frolov was supported by the Natural Sciences and
Engineering Research Council of Canada. The work of
A.L. Larsen was supported by the Danish Natural Science Research
Council under grant No. 11-1231-1SE

\end{document}